# On-chip single-mode thin film lithium niobate laser based on Sagnac loop reflectors

Shupeng Yu,[1,2] Zhiwei Fang,[3,7] Zhe Wang,[3] Yuan Zhou,[1,2] Qinfen Huang,[3] Jian Liu,[3] Rongbo Wu,[3] Haisu Zhang,[3] Min Wang,[3] and Ya Cheng[1,3,4,5,6,7,8]

[1]*State Key Laboratory of High Field Laser Physics and CAS Center for Excellence in Ultra-intense Laser Science, Shanghai Institute of Optics and Fine Mechanics (SIOM), Chinese Academy of Sciences (CAS), Shanghai 201800, China*
[2]*Center of Materials Science and Optoelectronics Engineering, University of Chinese Academy of Sciences, Beijing 100049, China*
[3]*The Extreme Optoelectromechanics Laboratory (XXL), School of Physics and Electronic Science, East China Normal University, Shanghai 200241, China*
[4]*Collaborative Innovation Center of Extreme Optics, Shanxi University, Taiyuan 030006, China.*
[5]*Collaborative Innovation Center of Light Manipulations and Applications, Shandong Normal University, Jinan 250358, People's Republic of China*
[6]*Hefei National Laboratory, Hefei 230088, China*
[7]*zwfang@phy.ecnu.edu.cn*
[8]*ya.cheng@siom.ac.cn*

**We demonstrate an on-chip single-mode Er$^{3+}$-doped thin film lithium niobate (Er: TFLN) laser which consists of a Fabry-Pérot (FP) resonator based on Sagnac loop reflectors (SLRs). The fabricated Er: TFLN laser has a footprint of 6.5 mm × 1.5 mm with a loaded quality (Q) factor of 1.6×10$^5$ and a free spectral range (FSR) of 63 pm. We generate the single-mode laser around 1550-nm wavelength with a maximum output power of 44.7 µW and a slope efficiency of 0.18 %.**

The thin film lithium niobate (TFLN) on insulator has emerged as a promising material platform for high-performance integrated photonic devices for both classical and quantum photonics applications due to its wide transparency window, high refractive index, as well as large acusto-optic, electro-optic, thermal-optic, and nonlinear optical coefficients [1–4]. In addition, doping the TFLN with rare earth ions offers a compatible approach to the realization of high-performance on-chip light sources of high coherence and wavelength tunability using the same fabrication technique developed for the passive TFLN devices [5, 6]. Indeed, the capability of creating ultra-low loss waveguides with meter-scale lengths wrapped within a footprint of centimeter-scale size on the rare earth ions doped TLFN provides the opportunity to produce miniaturized on-chip lasers of which the performances can approach that of fiber lasers in terms of output power, linewidth, and wavelength tunability while the production cost and the device size can both be reduced [7, 8]. Toward this goal, particularly for achieving the high output power, it is necessary to find a way to enable constructing long laser cavities with large gain volume in which the high energy can be stored, meanwhile, the cavity modes can be independently manipulated using end mirrors of sophisticated spectral band structures. So far, all of the rare earth ions doped TFLN lasers are demonstrated with the whispering gallery mode (WGM) resonators including microdisk and microring resonators, offering a large number of longitudinal modes with high-Q factors [9-24]. Thus, scaling up the output power from WGM TFLN lasers to compete with the conventional fiber laser seems unlikely to realize. However, in comparison with the WGM lasers, the Fabry-Pérot (FP) resonator-based lasers allow for the generation of high output power with the extended gain length in the resonator. The end reflectors are an essential component for FP resonators. For the integrated FP lasers without discrete components, the reflectors frequently chosen are distributed Bragg reflectors (DBRs) and high-reflectivity (HR) coatings [25-31]. Although the DFB reflectors are of ultra-small sizes and compatible with lithographic technology, it requires a very high fabrication resolution particularly for the high-index lithium niobate. Furthermore, the HR coatings require complicated fabrication processes, demanding multilayer films which are coated on the chip facets. Actually, due to the maturity in the fabrication technology of high-performance waveguide couplers on the TFLN substrate, the ideal reflectors for a monolithically integrated TFLN FP laser are the Sagnac loop reflectors (SLRs), each of which consists of a 2×2 3-dB optical coupler whose outputs are fused to form a loop waveguide. The coupler in the SLRs imposes a $\pi/2$ phase shift between the two outputs and the power coupling ratio is 50% [32]. Therefore, the device behaves as a perfect reflecting mirror and almost all of light is returned to the input waveguide. Notably, the SLRs can be readily fabricated using the photolithography assisted chemo-mechanical etching (PLACE), which has shown to enable scaling up the PICs on the TFLN platform due to the high fabrication efficiency, the sufficient fabrication resolution and the large footprint allowed by a single continuous fabrication process.

In this work, we demonstrate an on-chip single-mode FP resonator laser monolithically integrated on an Er[3+]-doped TFLN (Er: TFLN) platform. The laser cavity of a length of 5.8 mm is arranged between two integrated SLRs, thus the fabricated Er[3+]-doped TFLN FP laser has a footprint of only 6.5 mm × 1.5 mm. We achieve a maximum output power of 44.7 μW from two output ports. More importantly, for the first time to our best knowledge, a slope efficiency of 0.18% has been measured in our experiment, which is orders of magnitude higher than the TFLN WGM lasers reported so far [18-24]. The result shows the potential of generating high power on-chip TFLN lasers with decent slope efficiencies by optimizing the design parameters and fabrication qualities in the future.

Figure 1(a) shows the top view of the on-chip Er: TFLN FP resonator, which is fabricated on a 500-nm-thick X-cut Er: TFLN using the PLACE technology. The fabrication resolution of PLACE is ~200 nm, which is sufficient for fabricating waveguide-based photonic structures such as waveguide directional couplers and beam splitters. More fabrication detail can be found in Ref. 23. The doping concentration of erbium ions in TFLN is 0.5 mol%. The on-chip Er: TFLN FP resonator has a footprint of 6.5 mm × 1.5 mm, which consists of two SLRs and a straight Er: TFLN waveguide. The SLR has a footprint of 1.8 mm × 0.4 mm, which is produced by a 2×2 3-dB directional coupler of which the two output ends are fused to form a loop. The gain region is the straight waveguide between the two SLRs which is a length of ~5.8 mm. As illustrated at the bottom right corner of Figure 1 (a), the FP resonator is aligned in parallel to the Y crystallographic axis of the TFLN single crystal, as such a design can enable direct phase modulation on the output laser beam using monolithically integrated microelectrodes which generate electric fields along Z-axis to ensure the maximum electro-optical tuning efficiency. As shown in Figure 1(b) and (c), the length of the coupling region is ~300 μm, and the gap width is ~3.3 μm, respectively. As shown in Figure 1(d), the loop is designed to follow the Bezier curves for the minimization of the bending loss. The top width of our ridge waveguide on the TFLN is ~1 μm, and the etching depth of the ridge is ~210 nm, respectively.

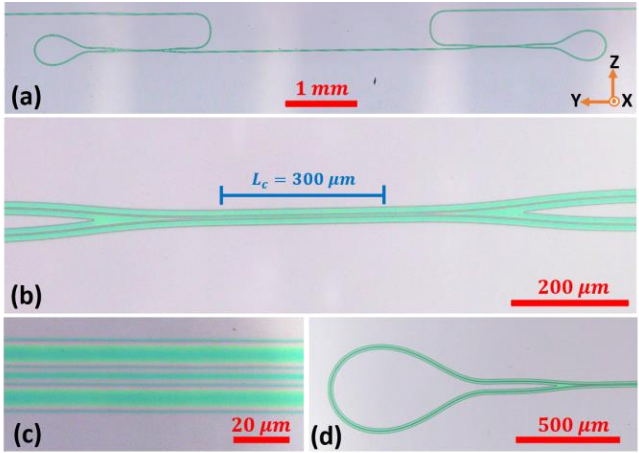

Fig. 1. (a) Optical microscope image of a Er: TFLN FP resonator (the arrows in bottom right corner illustrate the LN crystallographic axes X, Y, and Z). Zoomed-in optical microscope image of (b) an 3-dB directional coupler, (c) the coupling region, and (d) a Sagnac loop reflector.

waveguides. In Figure 2(a), we show the measured coupling ratio of the directional coupler with a fixed coupling length $L_c$ of 300 μm as a function of the gap width between the two waveguides at 1550 nm. Insets of Figure 2(a) show the simulated field profiles of the directional coupler for selected gap values. It can be seen that when the gap width of directional coupler is 3.3 μm, the coupling ratio is about 48% which is close to a 2×2 3 dB optical coupler. We then can calculate the normalized transmittance of the Sagnac loop $T_N$ using the following equation:

$$T_N = (K - (1-K))^2 \qquad [1]$$

where the $K$ is the coupling ratio of the coupler which is experimentally determined by the results in Figure 2(a). As shown in Figure 2(b), the normalized transmittance of the SLR with the 3.3-μm-width gap is about -28 dB at 1550 nm, indicating that most of the light entering the Sagnac loop will be reflected.

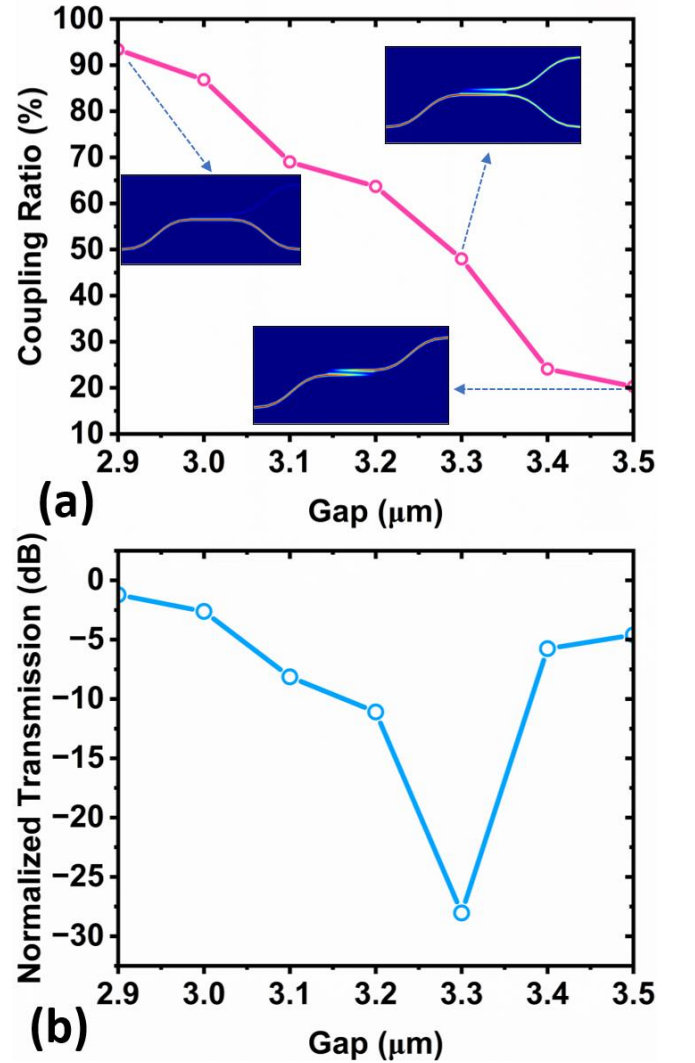

Fig. 2 (a) Experiment results of directional coupler's coupling ratio at 1550 nm with different gap. Insets : simulated field profiles for selected gap values. (b)Normalized transimission of Sagnac loop reflectors at 1550nm with different gaps according to the coupling ratios.

The reflectivity of the SLR depends on the coupling ratio of the directional coupler. To obtain the desirable reflectivity of our Saganc loop, we need the knowledge on the dependence of the coupling strength of the directional coupler on gap width between the two

The fabricated Er: TFLN FP resonators are characterized by spectroscopic measurement of the normalized optical transmission spectrum and laser emission. Light from a continuously tunable laser (CTL 1550, TOPTICA Photonics Inc.) was sent over a single-mode fiber into the Er: TFLN FP resonator via a lensed fiber, with the polarization state controlled by a 3-paddle fiber polarization controller (FPC561, Thorlabs Inc.). The transmitted light coupled out of the Er: TFLN FP resonator was collected again by a lensed fiber and directed into a photodetector (New focus 1811, Newport Inc.). As shown in Figure 3(b), a measured normalized transmission spectrum demonstrated that the Er: TFLN FP resonator has a loaded Q factor of $1.6 \times 10^5$ and an FSR of 63 pm. The Q factor of our Er: TFLN FP resonator is one order of magnitude higher than the TFLN FP resonator based on DBRs [25, 26]. Here, bidirectional pumping scheme was employed, so the pump beam from two laser diodes (CM97-1000-76PM, II-VI Laser Inc.) were coupled into the Er: TFLN FP resonator from both sides using the lensed fibers. The pump and signal light beams were combined and separated using fiber-based 980/1550 wavelength division multiplexers (WDM) at the input and output ports of the Er: TFLN FP resonator. Inset of Figure 3(a) shows the upconverted green fluorescence emission of the Er: TFLN FP resonator pumped by the 980-nm LD. An optical spectrum analyzer (OSA: AQ6375B, YOKOGAWA Inc.) was used to analyze the spectrum of the laser signal. As shown in Figure 3(c), only a single laser mode peak at 1552 nm wavelength can be observed in the spectrum recorded by the OSA in the wavelength range of 1500 nm - 1600 nm due to mode-dependent loss and gain competition [23]. The mode of the output laser was also imaged using an objective together with an infrared camera (InGaAs Camera C12741-03, Hamamatsu Photonics Co., Ltd.), as shown in the inset of Figure 3(c). One can clearly see that the generated laser is in the fundamental spatial mode. Therefore, the Er: TFLN FP resonator laser operates in not only single-longitudinal mode but also single-transverse mode around 1550-nm wavelength, which is highly desirable for numerous applications in the photonics community.

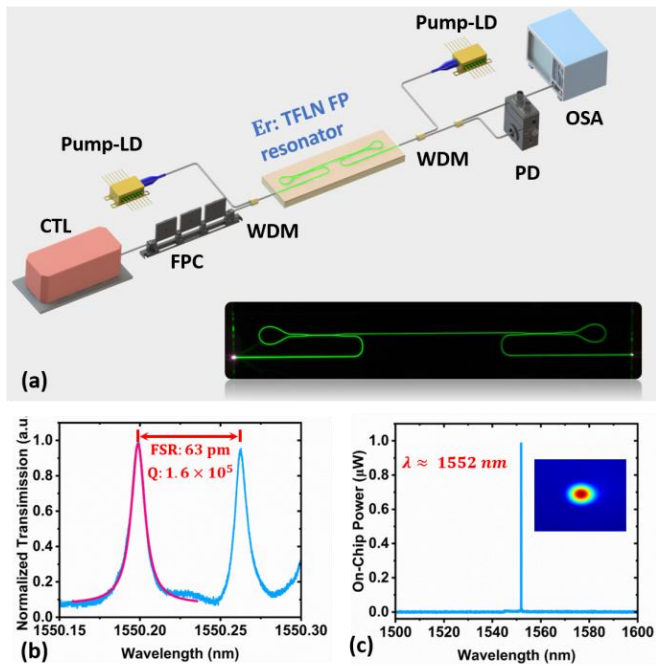

Fig. 3 (a) A schematic of the measurement setup to characterize the optical transmission and lasing emission of the fabricated Er: TFLN FP resonator laser, the lower-right inset : the green upconversion fluorescence of Er: TFLN FP resonator pumped by the 980-nm LD. (b) Normalized transmission spectrum of a Er: TFLN FP resonator, an optical resonance at 1550 nm is fitted with a Lorentzian line shape (red). (c) The spectrum of laser emission at 1552 nm, inset: The infrared image of the output port of the Er: TFLN FP resonator. (CTL: continuously tunable laser; FPC: fiber polarization controller; LD: laser diode; WDM: wavelength division multiplexers; OSA: optical spectrum analyzer; PD: photodetector).

Figure 4(a) highlights the single-mode spectrum of the fabricated Er: TFLN FP laser emission centered at λ ≈ 1544 nm, exhibiting a side-mode suppression ratio (SMSR) > 25 dB. Figure 4(b) shows the zoom-in spectrum of the laser emission, featuring a linewidth of 28 pm at 1544 nm. It is noteworthy that the measured 28 pm linewidth is limited by the resolving power of the OSA used in our experiment, which has a spectral resolution of ~0.01 nm. In Figure 4(c), the laser output power is plotted against the on-chip pump power coupled into the Er: TFLN FP resonator. The lasing threshold was measured to be ~6 mW by linear fitting, and we achieved a slope efficiency of 0.18 %, far higher than the previously reported slope efficiencies in the Er: TFLN lasers [18-24]. As shown in Figure 4(d), we measured a maximum output power of 44.6 μW for the laser operation at 1544 nm wavelength. However, we clarify here that the output power measured in Figure 4(d) is the total output powers generated from the SLR-based Er: TFLN FP laser by summing up the powers measured at the two output ports.

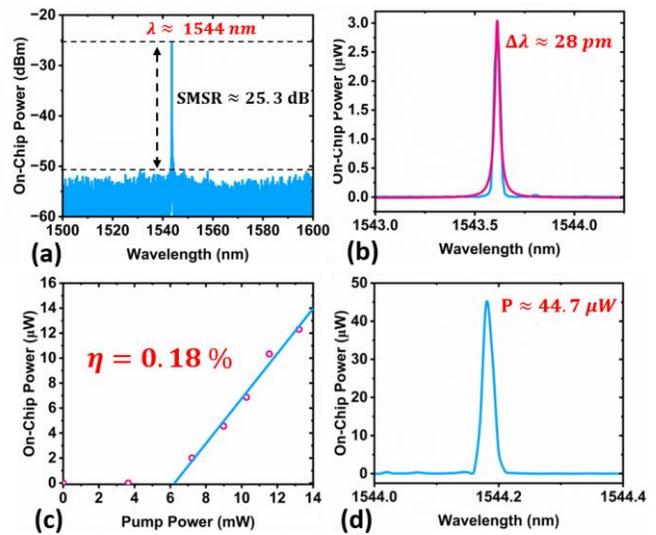

Fig. 4. (a) Output optical spectrum of the Er: TFLN FP resonator laser from 1500 nm to 1600 nm. (b) The enlarged spectrum around wavelength 1544 nm, the lasing peak is fitted with a Lorentzian line shape (red). (c) On-chip laser power of Er: TFLN FP resonator laser vs absorbed pump power. (d) Spectral power density of Er: TFLN FP resonator laser emission.

In conclusion, we demonstrate an on-chip single-mode FP resonator laser monolithically integrated on the Er: TFLN platform. These Er: TFLN FP lasers can efficiently generate single-mode laser emissions around 1550-nm wavelength when pumped by 980-nm LD. The lasing threshold and the slope efficiency are measured to be ~6 mW and 0.18 %, respectively. The maximum output power of our Er: TFLN FP resonator laser is ~44.7 μW. We envisage that the result reported here will open the avenue for high-power single-mode electro-optical tunable waveguide lasers of miniaturized footprint as well as compatibility of monolithic integration with other existing passive and active photonic structures on the TFLN platform.

**Funding.** National Key R&D Program of China (2019YFA0705000), National Natural Science Foundation of China (Grant Nos. 12274133, 12004116, 12104159, 12192251, 11933005, 12134001, 61991444, 12204176), Science and Technology Commission of Shanghai Municipality (NO.21DZ1101500), Shanghai Sailing Program (21YF1410400), Shanghai Pujiang Program (21PJ1403300).